%% file: main.tex
\documentclass[10pt,twocolumn,letterpaper]{article}

\usepackage{cvpr}

\input{preamble}

\definecolor{cvprblue}{rgb}{0.21,0.49,0.74}
\usepackage[pagebackref,breaklinks,colorlinks,allcolors=cvprblue]{hyperref}

\def\paperID{10476} 
\def\confName{CVPR}
\def\confYear{2025}

\title{\ApproachName: Generating Vector Displacement Maps From a Single Image}

\author{
\hspace{-2em}
Yuezhi Yang$^{1}$\thanks{work partially done when interning at Adobe.} \quad
Qimin Chen$^{2}$ \quad
Vladimir G. Kim$^{3}$ \quad
Siddhartha Chaudhuri$^{3}$ \\
Qixing Huang$^{1}$ \quad
Zhiqin Chen$^{3}$ \\
{
\textsuperscript{1}The University of Texas at Austin
\quad
\textsuperscript{2}Simon Fraser University
\quad
\textsuperscript{3}Adobe Research
}
}

\begin{document}
\maketitle

\input{sec/0_abstract}
\vspace{-4mm}
\input{sec/1_intro}

\input{sec/2_relatedwork}
\input{sec/3_method}

\input{sec/4_experiments}

\input{sec/5_conclusion}
{
\small
\bibliographystyle{ieeenat_fullname}
\bibliography{main}

}

\input{sec/X_suppl}

\end{document}

%% file: preamble.tex
\usepackage{color}
\usepackage{multirow}
\usepackage[accsupp]{axessibility}

\usepackage{bm}

\renewcommand{\paragraph}[1]{\vspace{\parskip}\noindent\textbf{#1}}

\definecolor{turquoise}{cmyk}{0.65,0,0.1,0.3}
\definecolor{purple}{rgb}{0.65,0,0.65}
\definecolor{dark_purple}{rgb}{0.5,0,0.5}
\definecolor{dark_green}{rgb}{0, 0.5, 0}
\definecolor{orange}{rgb}{0.8, 0.6, 0.2}
\definecolor{red}{rgb}{0.8, 0.2, 0.2}
\definecolor{darkred}{rgb}{0.6, 0.1, 0.05}
\definecolor{blueish}{rgb}{0.0, 0.3, .6}
\definecolor{light_gray}{rgb}{0.7, 0.7, .7}
\definecolor{pink}{rgb}{1, 0, 1}
\definecolor{greyblue}{rgb}{0.25, 0.25, 1}

\newcommand{\ApproachName}{GenVDM\xspace}

\newcommand{\SupplementaryMaterial}{{Supplementary Material}\xspace}

\DeclareMathOperator*{\argmin}{argmin}

%% file: sec/0_abstract.tex
\begin{abstract}
We introduce the first method for generating Vector Displacement Maps (VDMs): parameterized, detailed geometric stamps commonly used in 3D modeling. Given a single input image, our method first generates multi-view normal maps and then reconstructs a VDM from the normals via a novel reconstruction pipeline. We also propose an efficient algorithm for extracting VDMs from 3D objects, and present the first academic VDM dataset. Compared to existing 3D generative models focusing on complete shapes, we focus on generating parts that can be seamlessly attached to shape surfaces. The method gives artists rich control over adding geometric details to a 3D shape. Experiments demonstrate that our approach outperforms existing baselines. Generating VDMs offers additional benefits, such as using 2D image editing to customize and refine 3D details. Code and data are available at \href{https://yyuezhi.github.io/GenVDM/}{https://yyuezhi.github.io/GenVDM/}.
\vspace{-2mm}
\end{abstract}

%% file: sec/1_intro.tex
\section{Introduction}
\label{sec:intro}

Generative neural models for 3D shape synthesis is a rapidly advancing research area~\cite{DBLP:journals/corr/abs-2210-15663}. However, they are still not widely adopted in artistic workflows for two main reasons. First, synthesizing fine geometric details is challenging due to the heterogeneity of 3D representations and the lack of detailed 3D training data. Second, existing neural tools lack the precise spatial and compositional controls needed by 3D artists. To address these limitations, instead of reinventing the 3D modeling stack to accommodate generative AI, we draw inspiration from an existing workflow in which an artist starts with a base mesh and ``stamps’’ the desired details onto the 3D surface (see Figure~\ref{fig:teaser}). These smaller stamps are easier to generate than full-scale 3D models, fit seamlessly into existing workflows, eliminate artists' dependence on expensive and limited third-party stamp libraries, and provide full artistic control over spatial arrangement and composition.

We chose the {\em vector displacement map} or VDM as our stamp representation. A VDM assigns an arbitrary 3D displacement to every point in a 2D rectangle, warping the sheet to form a curved surface with complex geometric features, such as overhangs and cavities. It is widely supported in 3D software~\cite{Maya, ZBrush, Mudbox, Blender} and compactly stored as a vector field over a UV image domain. While using VDMs is commonplace, authoring them is extremely challenging, and artists usually depend on packs of VDMs created by third parties (analogous to brushes in digital painting tools), with limited customization or generality. Image or text-driven stamp generation could drastically expand the scope of VDM usage by providing artists with custom stamps on demand.

\input{figs/teaser}

In this paper, we propose the first neural pipeline to generate a VDM from a single RGB image. To achieve this, we address two main technical challenges. The first challenge is that existing generative models are not suitable for VDM generation: generating a 3D object usually does not also produce a parametric 2D domain for stamp application, and predicting a depth map from a single image does not capture complex high-amplitude variations, overhangs, and occlusions; see Figure~\ref{fig:qualitative_comparison}. Thus, we develop a three-step method. First, given an input RGB image (which can also be generated with existing text-to-image models), we predict normal maps from multiple viewing directions to resolve occlusions that may be hidden in a single view. Second, we reconstruct a mesh (which need not have disk topology) by fitting a neural SDF to the multi-view normal maps and polygonizing the result. Third, we use a neural deformation model to displace points on a 2D rectangle to fit the mesh, forming the final VDM.

The second challenge in training a generative VDM model is the absence of training data. We tackle it by building an interactive tool to segment interesting semantic and geometric regions from Objaverse 3D models~\cite{objaverse}, and then develop a geometry processing pipeline for converting these regions into a VDM representation, creating a dataset of 1,200 VDM patches used for training. Our pipeline is robust enough to analyze polygon soups in the wild, which we achieve by re-sampling the selected regions and reconstructing a single connected surface after removing outliers. We then deform the resulting mesh to obtain a co-planar boundary that can be seamlessly attached to a flat base tile over which the VDMs are typically defined. The processed shapes can then be rendered and used to finetune the multi-view normal generation model.

We compare our method to state-of-the-art shape generation techniques~\cite{wonder3d,magic123,LRM}, as well as to reconstructing a heightfield (i.e. a {\em scalar} displacement map) from estimated depth~\cite{depthanything}. We use a collection of images depicting parts commonly used in VDMs (e.g., facial elements, decorations), and evaluate using visual fidelity~\cite{Seitzer2020FID} and semantic similarity~\cite{CLIP} metrics. Our method outperforms others due to its ability to handle smaller VDM-like regions. Note also that other mesh generation methods do not produce a displacement map -- which can have both ``outward'' and ``inward'' displacements -- and thus their output can only be additively combined with the base shape, e.g., they are not able to introduce cavities like an eye or a mouth in Figure~\ref{fig:teaser}.

To summarize, our contributions are:
\begin{itemize}
    \item The first generative ML pipeline for VDMs;
    \item A robust method to reconstruct VDMs from multi-view normal maps produced by image diffusion models;
    \item A novel VDM extraction pipeline to efficiently extract and process patches from 3D objects to produce VDMs;
    \item The first public dataset of VDMs for academic research.
\end{itemize}

%% file: figs/teaser.tex
\begin{figure}[t]
  \centering
   \includegraphics[width=1.0\linewidth]{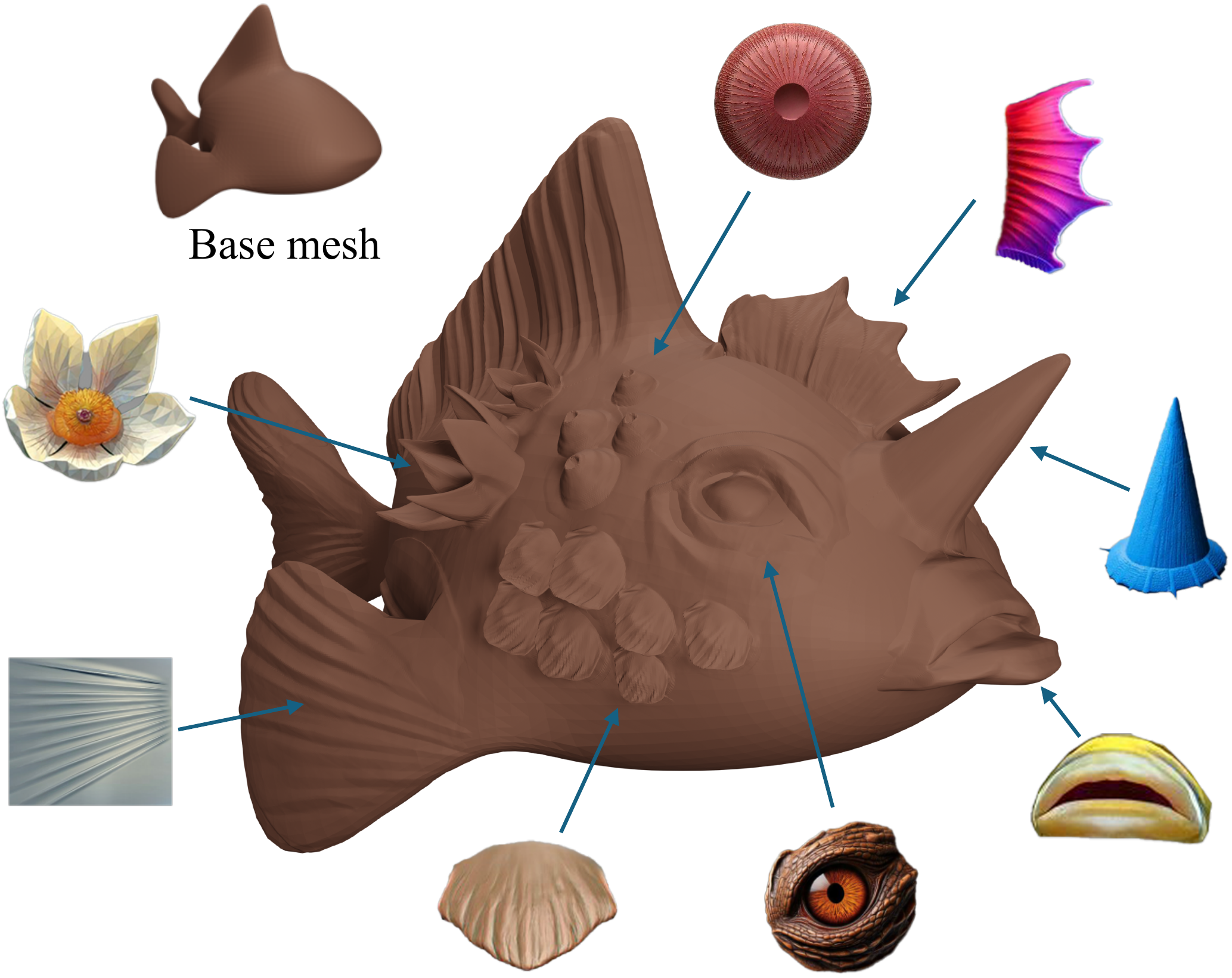}
   \caption{We introduce \ApproachName, a method that can generate a highly detailed Vector Displacement Map (VDM) from a single input image. The generated VDMs can be directly applied to mesh surfaces to create intricate geometric details. Note that the thumbnails represent plain 2D RGB image sources.}
   \label{fig:teaser}
\end{figure}

%% file: sec/2_relatedwork.tex
\section{Related work}
\label{sec:related}

\paragraph{Vector Displacement Maps.}
Texture mapping~\cite{texturemap,texturemapsurvey} is the dominant solution in the industry to add complex surface details to shapes without increasing mesh complexity. Accompanying it are many techniques that hallucinate complex geometric details, such as bump mapping~\cite{bumpmap}, horizon mapping~\cite{horizonmap}, and parallax mapping~\cite{parallaxmap}. Unlike those techniques that do not change the geometry of the shape, displacement mapping~\cite{displacementmap,displacementmapapply,displacementmapsurvey} adds geometric details by subdividing the original geometry into finer polygons and then displacing each vertex in its normal direction by a height value indexed from the displacement map (although some versions of displacement mapping can be done in the pixel space without changing the original geometry~\cite{viewdependent_displacementmap}).

While a displacement map can be considered as a single-channel image or heightfield, a vector displacement map (VDM) can be seen as a three-channel image, where each pixel contains a 3D displacement vector. VDMs naturally support representing more complex geometries with less distortion compared to displacement maps, and both are used in 3D modeling tools to create geometric details. Research on displacement maps and VDMs has focused on texture synthesis from examples~\cite{ying2001texture}, and synthesis of human body and face meshes for shape reconstruction~\cite{image2humanbody,image2face}. VDMs conceptually resemble Geometry Images~\cite{GeometryImages}, and some recent works adopt image diffusion models for generating Geometry Images to synthesize 3D shapes~\cite{GeometryImageDiffusion,AnotherGeometryImageDiffusion}. To our knowledge, there is no prior work on generative models of VDMs, nor a public research dataset for VDMs.

\input{figs/overview}

\paragraph{Image-to-3D.}
Early works on single-view 3D reconstruction~\cite{3d_r2n2,pixel2mesh,atlasnet,im_net,occnet,disn,pixelnerf} mostly adopt feed-forward neural networks trained on limited data~\cite{shapenet}. More recent work~\cite{point_E,shape_E,3dshape2vecset,clay} trained on large 3D datasets~\cite{objaverse} has shown significantly improved generalizability to novel shape categories. With the introduction of text-to-image diffusion models~\cite{latent_diffusion_models_SD, SDxl}, a line of work~\cite{realfusion,Make_it_3d} achieved zero-shot single-image-to-3D with score distillation sampling (SDS)~\cite{dreamfusion} by distilling 2D diffusion priors into 3D representations with per-shape optimization.

Another line of work~\cite{3DiM,zero123} utilizes image diffusion models for novel view synthesis conditioned on an input image and a relative camera pose. Such models produce images of the object from different views, therefore 3D object can be reconstructed by SDS-based optimization~\cite{magic123} or a feed-forward reconstruction network~\cite{One-2-3-45}. These methods inspired a series of subsequent work that finetunes pretrained image diffusion models to directly generate 3D-consistent multi-view images of target output shape given a single-view image, where output shape can be reconstructed from generated multi-view images via optimizing a neural field or mesh~\cite{SyncDreamer,zero123++,wonder3d}, a 3D diffusion reconstruction network~\cite{One-2-3-45++}, or a feed-forward large reconstruction model powered by Transformers~\cite{LRM, Instant3D, PF_LRM, Lgm, Gs_lrm, LRM_Zero, Instantmesh, CRM, mvd2}. Most recently, image diffusion models have been replaced by video diffusion models to achieve better 3D consistency of generated views~\cite{Vfusion3d, Sv3d}.

\paragraph{Modeling by Parts.}
The use of small building components to compose complex shapes has been widely studied in modeling-by-assembly systems~\cite{ModelingByExample,Shuffler}. Before generative AI rose to prominence, these systems relied on part databases~\cite{SidVangelisAssembly} (or shape databases from which parts could be cut out), and focused on building tools to help users find the right parts~\cite{sketch2design_cgf13,akzm_shapeSynth_eg14,AttribIt,PartBasedStructureRecovery} and assemble them meaningfully~\cite{PartBasedRecombination,FitAndDiverse,ComplementMe}. As a variation, methods were developed to extract and transfer detailed patches from a shape to another~\cite{Takayama2011GeoBrush}. A few papers studied joint synthesis and layout of parts~\cite{Li2017GRASS}, but the synthesis was conditioned only on the layout and not on user input, and the focus was on whole-shape generation and not adding detail to existing ones.

Relying on existing part datasets or part generation without user control, and on complex, non-standard, topology-sensitive mesh fusion algorithms limits the utility of these older methods. Our approach generates detailed complementary geometry in-situ from the image prompt, and our generated VDMs are defined over parameterized 2D domains which are suitable for seamlessly blending onto 3D models, with industry-wide support.

\vspace{-1mm}

%% file: figs/overview.tex
\begin{figure*}[t]
  \centering
   \includegraphics[width=1.0\linewidth]{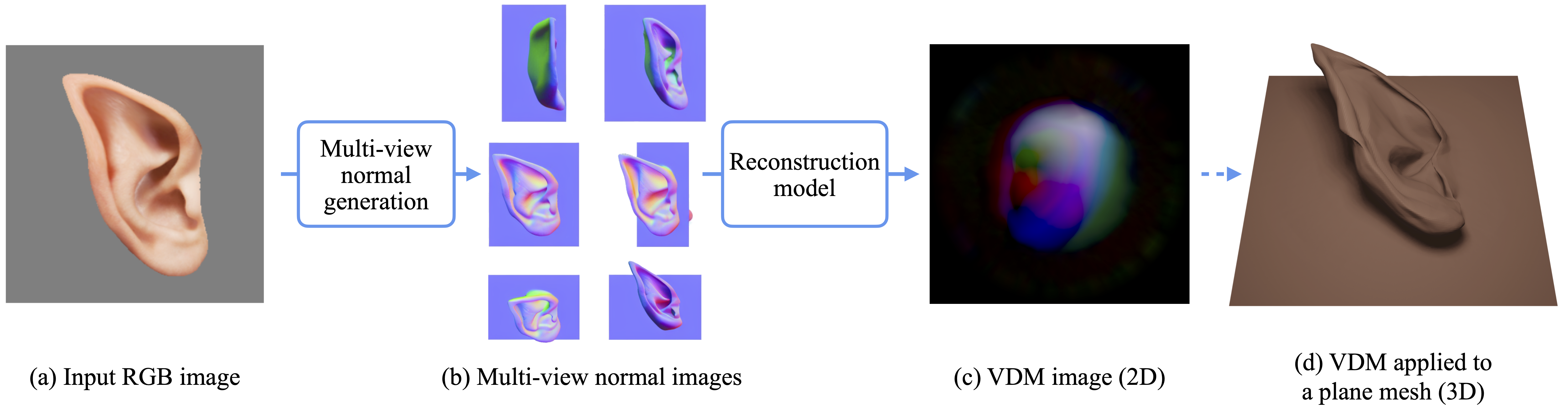}
   \caption{Overview of our image-to-VDM pipeline. Given an input image, we first add a gray square behind the object/part in the image as background, so the image resembles a textured VDM applied to a square mesh, as in (a). Then we utilize a multi-view image diffusion model to generate six normal maps with pre-defined camera poses, as in (b). The multi-view normal maps effectively represent the geometry of the VDM when applied to a square mesh, and thus we can reconstruct the VDM from these normal maps, as in (c). The reconstructed VDM can then be applied to various surfaces as in (d).}
   \label{fig:overview}
\end{figure*}

%% file: sec/3_method.tex
\section{Method}
\label{sec:method}
\vspace{-2mm}
Our image-to-VDM pipeline is shown in Figure~\ref{fig:overview}. Similar to other methods in the literature, we follow an approach that first generates multi-view images of the target object with an image diffusion model and then reconstructs the object from the generated images. In particular, we only generate normal maps of the object as we are only interested in the geometric details. Details of the multi-view normal generation are described in Section~\ref{sec:multiview-normal-gen}. Next, we reconstruct the VDM from the multi-view normals. As VDMs have specific properties and constraints, reconstructing them is highly non-trivial. We report our attempts and solutions in Section~\ref{sec:vdm-recon}. Finally, as there is no publicly available dataset for VDMs, we designed an efficient tool for extracting shape patches from Objaverse~\cite{objaverse}, and devised algorithms to process those patches for use as training data. We describe the data processing pipeline in Section~\ref{sec:data-preparation}.

\vspace{-3mm}

\subsection{Multi-View Normal Map Generation}
\label{sec:multiview-normal-gen}
\vspace{-3mm}
We opt to finetune an image diffusion model to generate multi-view images, as the pretrained image diffusion model offers strong generalizability. As will be shown in our experiments, our model, trained on a small dataset of 1,200 examples, works on a large variety of shapes.

Specifically, we adopt Zero123++~\cite{zero123++} as the backbone for our multi-view diffusion model. Zero123++ is an image-to-multiview model based on Stable Diffusion~\cite{latent_diffusion_models_SD}. Given an input image, Zero123++ generates a $960 \times 640$ image representing six multi-view images in a $3 \times 2$ grid, where the six images have pre-defined camera poses so they can be easily used for 3D reconstruction. However, the pre-defined camera poses in Zero123++ fully surround the object, e.g., there are front views and back views of the object. In our pipeline, since we are aiming to generate VDMs, the back views of the object are unnecessary. Therefore, we re-designed the camera poses of the six images. As shown in Figure~\ref{fig:overview} (b), assuming that the front view (see (a) for an example) has $($elevation angle, azimuth angle$) = (0^{\circ},0^{\circ})$, we define the six camera poses as $(0^{\circ},-60^{\circ})$, $(0^{\circ},-30^{\circ})$, $(0^{\circ},30^{\circ})$, $(0^{\circ},60^{\circ})$, $(45^{\circ},0^{\circ})$, $(-45^{\circ},0^{\circ})$. We also adopt orthographic cameras to reduce distortion, and let the model generate a normal map of the object for each camera pose.
To train the model, we render single-view RGB images as input and multi-view normal maps as ground truth output. Details about training data are described in Section~\ref{sec:data-preparation}. Note that the input image does not have to be a front view; we render random views for training so the model can handle images from various viewpoints. We fine-tuned the checkpoint provided by Zero123++~\cite{zero123++} on 8 NVIDIA A100 GPUs for 3 days.

\subsection{VDM Reconstruction}
\label{sec:vdm-recon}

\input{figs/recon}

Reconstructing 3D shapes from multi-view images has been well studied in the text/image-to-3D literature. Most recent methods adopt a feed-forward large reconstruction model (LRM) to directly generate a 3D shape from multiple input images of different viewpoints~\cite{LRM, Instant3D, PF_LRM, Lgm, Gs_lrm, meshlrm}. Therefore, a straightforward way for reconstructing VDMs is to train a similar LRM to take the normal maps as input and directly regress a VDM image. 
However, given limited VDM training shapes, 

our LRM trained on a small dataset is unlikely to generalize as well as other LRM models trained on larger datasets,
therefore leading to suboptimal results.

Given the above discussions, we adopt a slower but more robust per-shape optimization approach. Given the six normal maps with pre-defined fixed camera poses, we want to optimize a 3D representation to converge to the target 3D shape with supervision provided by differentiable rendering. A naive approach would be to initialize with a discretized square mesh and optimize with mesh-based differentiable rendering. However, as has been shown in other methods~\cite{Large_steps_in_inverse_rendering, nvdiffrast}, differentiable rendering on meshes is often problematic and requires careful design of regularization losses and tuning of hyperparameters. As we will show later, even with ground truth 3D supervision, optimizing a discretized mesh to fit the target shape is not an easy task.

Therefore, we devise a two-step approach, as shown in Figure~\ref{fig:recon}, to first optimize a neural SDF field to reconstruct a 3D shape from the multi-view normal maps, and then parameterize the 3D shape into a VDM image. We utilize the method proposed in Wonder3D~\cite{wonder3d} for the first step, with the only modification being that we removed $L_{rgb}$, the loss term to punish the difference between rendered RGB images and the ground truth, as we do not predict multi-view RGB images. Since we always put a grey square as background in our input images, the shape we obtained via optimization has a solid plane-like primitive where the object/part is attached to, see Figure~\ref{fig:recon} (b); then we can extract a mesh from the neural SDF field and easily separate a single layer of mesh that represents the VDM.

\input{figs/vdm_fitting}
\input{figs/data}

The next step is to parameterize the mesh into a VDM image. Since the mesh is reconstructed from sparse-view images, its geometry is often noisy and riddled with small holes and large gaps, see Figure~\ref{fig:vdm_fitting} (a) left. To convert it into a VDM, we will need to fix its topology so that it is topologically equivalent to a plane; and then we will apply a mesh parametrization method to obtain its Tutte embedding on a square, so that each pixel on the square can be assigned with a displacement vector. However, as shown in Figure~\ref{fig:vdm_fitting} (a), although the state-of-the-art topology fixing algorithms \cite{ToCutOrToFill} can fix the topology, the result is often not satisfactory, e.g., a gap that should have been filled is being cut, see Figure~\ref{fig:vdm_fitting} (a) middle where the helix of the ear is cut in half. As a result, after applying \cite{sheffer2005abf} to obtain its embedding on a plane, we see large distortions and noise in the final VDM, see Figure~\ref{fig:vdm_fitting} (a) right where the upper part of the ear is missing due to distortion.

An alternative is to initialize with an optimizable square mesh and optimize it using a reconstruction loss with respect to the target mesh, 
as shown in Figure~\ref{fig:vdm_fitting} (b). However, as mentioned, it is often required to have carefully designed regularization losses when a mesh is to be optimized. When adopting a naive optimization method proposed in \cite{chen2024textguided}, the resulting mesh exhibits large distortion.

Therefore, instead of tuning the mesh optimization algorithm, inspired by AtlasNet~\cite{atlasnet} and Deep Geometric Prior~\cite{Deep_geometric_prior}, we propose to deform the square mesh with a neural deformation field parameterized by a Multilayer Perceptron (MLP). The MLP acts as a natural regularizer, as its inductive smoothness bias encourages smoothness of the deformation. We define the square as $\{p \; | \; p \in [0,1]^2\}$, and the MLP $\phi_{\theta}$ with optimizable parameters $\theta$. Then, given any 2D point $p$ in the square, we obtain its corresponding 3D point $p' = \phi_{\theta}(p)$ in the deformed shape. Therefore, for each optimization step, we sample a grid of 2D points in $[0,1]^2$, apply $\phi_{\theta}$ to obtain the deformed 3D points, and then compute the symmetric Chamfer Distance between the deformed 3D points and the ground truth points sampled from the target mesh. We also include a loss to maintain the square boundary. Therefore our optimization objective is
\begin{equation}
\resizebox{0.9\linewidth}{!}{
$\begin{split}
    \argmin_{\theta} \;\; \mathbb{E}_{P,Q} \; & \frac{1}{|P|}\sum_{p \in P}\min_{q \in Q}\|\phi_{\theta}(p)-q\|_2^2 +
    \frac{1}{|Q|}\sum_{q \in Q}\min_{p \in P}\|\phi_{\theta}(p)-q\|_2^2 + \\
    & \frac{1}{|\partial P|}\sum_{p \in \partial P} \|\phi_{\theta}(p) - \text{proj}(p)\|_2^2,
\end{split}$}
\end{equation}

where $P$ and $Q$ are sets of sampled points from $[0,1]^2$ and the target mesh, respectively. $\partial P$ contains all the boundary points in $P$ and $\text{proj}(p)$ maps $p$ to a corresponding 3D point in a pre-defined square boundary.
After optimization, we can sample a regular grid of points in $[0,1]^2$ and compute their 3D displacement vectors from $\phi_{\theta}$ to obtain the VDM image, as shown in Figure~\ref{fig:vdm_fitting} (c).

\subsection{Data Preparation}
\label{sec:data-preparation}

To the best of our knowledge, there is no publicly available dataset for VDMs. Therefore, we developed a data processing pipeline so we can efficiently annotate interesting parts from objects and then convert the parts into VDMs. In fact, our data processing pipeline does not produce true VDMs, but rather, shapes that look like VDMs, which are good enough to train our multiview generation model, see Figure~\ref{fig:data}. If needed, our VDM reconstruction method in Section~\ref{sec:vdm-recon} can be used to obtain readily usable VDMs.

To construct our VDM training dataset, we crop parts from the Objaverse~\cite{objaverse} dataset. We first create a keyword filtering list and apply the filter on Objaverse shape captions~\cite{luoCAP3D2023scalable,luoCAP3D2024view}. As VDMs are mostly used to model organic parts, we select objects likely to contain such parts, e.g., animals and characters.

We then developed a UI to precisely crop a part from a 3D object. This is achieved by a 3D lasso tool, where the user only needs to select a ring of points along the cutting boundary of the desired part. Our algorithm connects the points to form a cut and extracts the part from the object. Note that the part may not be a single connected mesh -- it may comprise several sub-meshes. Hence, we remesh the part into a single connected mesh. We first densely sample points on the part, and then remove interior points by computing winding numbers~\cite{FastWindingNumber}. For the remaining points, we performed screen Poisson surface reconstruction~\cite{Screened_poisson} to obtain a single connected mesh (Figure~\ref{fig:data} (b)). Our 3D lasso tool has proven to be quite efficient. Annotating our entire dataset with 1,200 parts took only 24 man-hours.

After obtaining the parts, we will then stitch each part to a square mesh to mimic the appearance of a VDM applied to a plane. Note that in almost all cases, the vertices on the boundary of each part are not coplanar, therefore, additional steps are required to make them coplanar. We first determine the plane via least squares plane fitting with respect to the boundary vertices. Then we project the boundary vertices to the plane, and adopt a method similar to Poisson Image Editing~\cite{Poisson_image_editing} to deform the part so that it follows the new coplanar boundary. Denote the set of all boundary vertices in the part (before projection) as $B$ and non-boundary vertices as $A$; also denote the set of all edges as $E$. Denote the coplanar boundary vertices after projection as $B'$, and the non-boundary vertices after deformation as $A'$. For each point $p$ in $A$ or $B$, denote its corresponding point in $A'$ or $B'$ as $p'$. Then our new vertices after mesh deformation can be obtained by solving a quadratic error function
\begin{equation}
\scalebox{0.9}{$
    \argmin_{A'} \;\; \mathbb{E}_{(p,q) \in E} \; \|(p'-q')-(p-q)\|_2^2.
$}
\end{equation}

The minimization objective is to ensure that the gradients on the mesh are preserved after deformation, while the target coplanar boundary points $B'$ are also strictly followed.

We then place the deformed part on a square mesh so that the boundary vertices and the square mesh vertices are coplanar. Once the part is attached to the square mesh, we perform one additional Laplacian smoothing step on the vertices close to the boundary to remove boundary noise, see Figure~\ref{fig:data} (c). 
We always keep the square mesh gray and assign a random color to the part. We also perform translation, scaling, and rotation augmentation to the part to enrich the diversity of the dataset. Finally, for each shape, we render several RGB images from different viewpoints to serve as the training input to the multi-view normal generation model, and six normal maps in pre-defined camera poses as the ground truth output, see Figure~\ref{fig:data} (d, e).

%% file: figs/recon.tex
\begin{figure*}[t]
  \centering
   \includegraphics[width=1.0\linewidth]{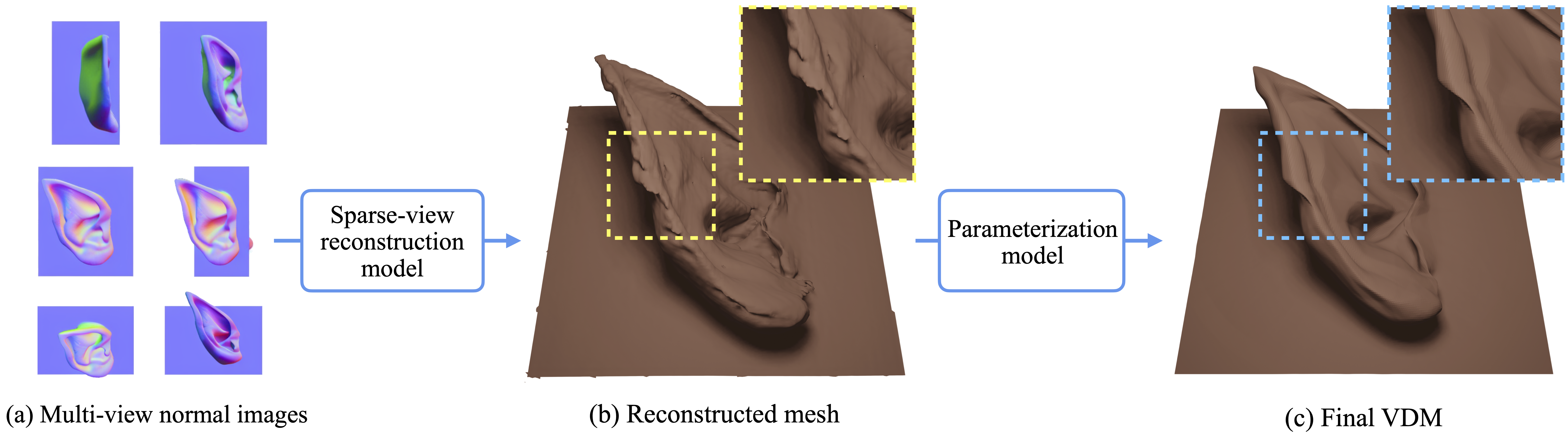}
   \caption{Reconstructing VDM from multi-view normal maps. We adopt a two-step approach. First, we reconstruct an accurate (but perhaps noisy) mesh (b) from the multi-view normals (a) with differentiable rendering and neural SDF representation. Then we parameterize the mesh by fitting a deformable square to it with a neural deformation field, as in (c). An VDM image can thus be obtained by discretizing the square into pixels and infer each pixel's displacement from the neural deformation field. The whole reconstruction pipeline takes about 6 minutes for each shape on an NVIDIA A100 GPU, where each step takes about 3 minutes.}
   \label{fig:recon}
\end{figure*}

%% file: figs/vdm_fitting.tex
\begin{figure}[t]
  \centering
   \includegraphics[width=1.0\linewidth]{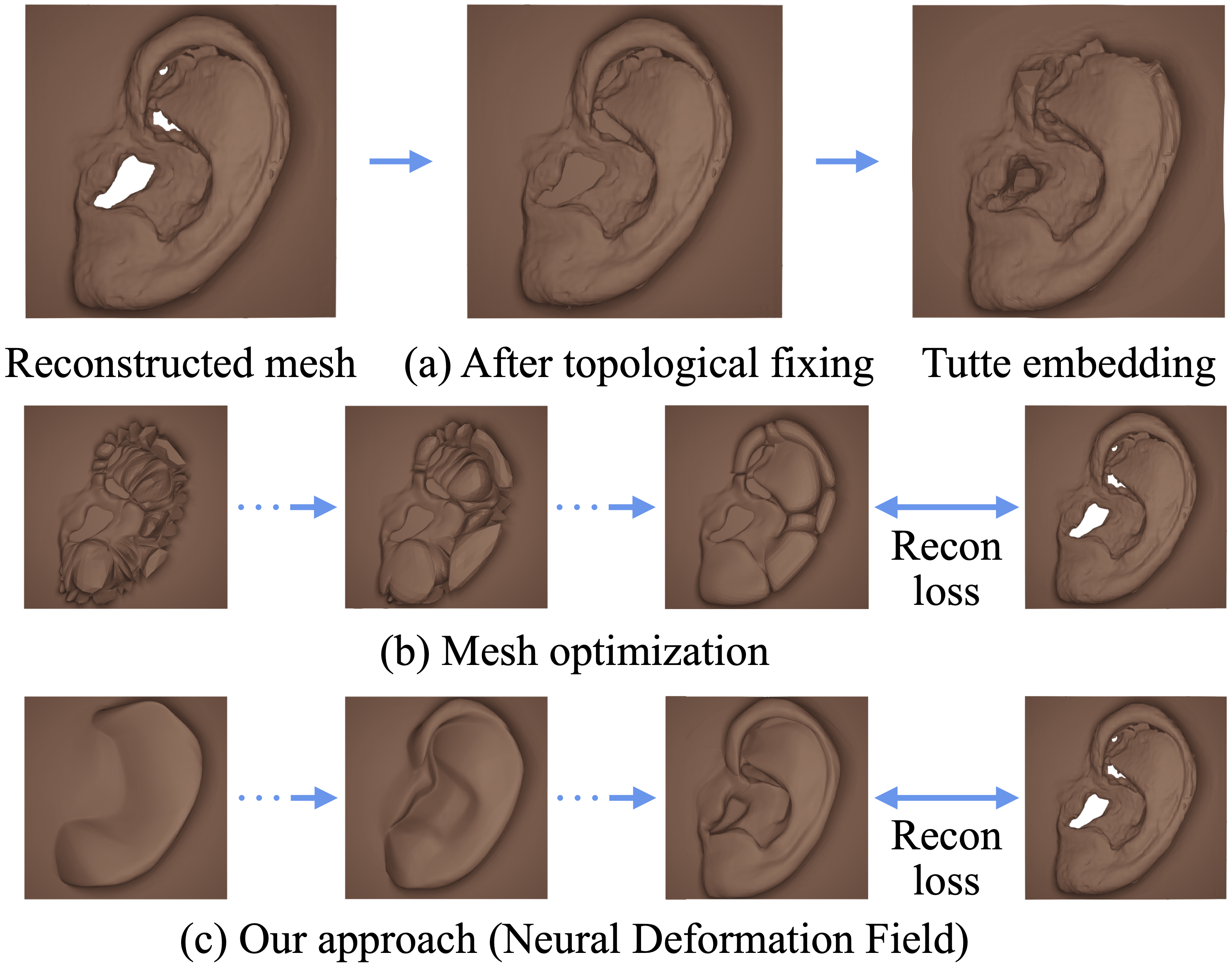}
   \caption{Comparison of different approaches for parameterizing a shape into VDM. (a) Topology fixing and Tutte embedding with classic tools leads to noise and distortion. (b) Fitting a plane mesh to the target mesh leads to large distortion. (c) Our approach by applying a neural deformation field to a parametric square leads to clean and high-quality reconstruction.}
   \label{fig:vdm_fitting}
\end{figure}

%% file: figs/data.tex
\begin{figure*}[t]
  \centering
   \includegraphics[width=0.95\linewidth]{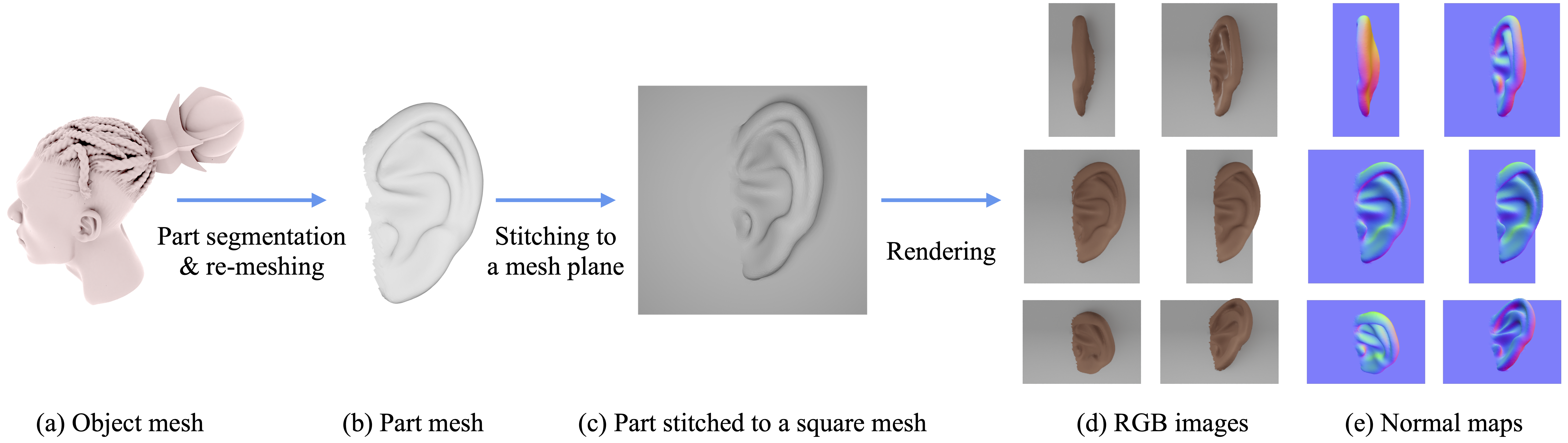}
   \caption{Data preparation. For each interesting object (a), we use a 3D lasso tool to segment out interesting parts. For each part, we densely sample points on the part's surface and then perform Screened Poisson Surface Reconstruction~\cite{Screened_poisson} to obtain a single connected mesh (b). We then stitch the mesh to a square mesh with an algorithm inspired by Poisson Image Editing~\cite{Poisson_image_editing} (c). Afterwards, we can color the part and render RGB images (d) and normal maps (e) for training the image diffusion model.}
   \label{fig:data}
\end{figure*}

%% file: sec/4_experiments.tex
\section{Experiments}
\label{sec:exps}

In this section, we verify the effectiveness of our method by comparing it with various state-of-the-art methods. We also validate our design choices in ablation studies. Finally, we present additional results produced by our method, show applications of VDMs on adding details to geometry, and demonstrate how users can customize VDMs by simply editing the input images. We will make our code, trained model weights, and dataset available to the public.

\subsection{Vector Displacement Map Generation}
\label{sec:exp-comparison}

\paragraph{Baselines.}
Since there is no prior work on generating VDMs from single view images, we compare our method with methods that perform a similar task, namely, single-view image to 3D reconstruction. Specifically, we compare our method with Wonder3D~\cite{wonder3d}, Magic123~\cite{magic123}, Large Reconstruction Model (LRM)~\cite{LRM}, as well as a \textit{scalar} displacement map (scalar DM) reconstruction method based on DepthAnything~\cite{depthanything}.
Given an input image, Wonder3D~\cite{wonder3d} generates multi-view RGB and normal images and optimizes a neural SDF field to reconstruct the 3D shape from the multi-view images.
Magic123~\cite{magic123} uses the SDS loss~\cite{dreamfusion} to optimize the 3D shape while applying a reconstruction loss on the input view.
LRM~\cite{LRM} generates multi-view RGB images and trains a Transformer-based feed-forward model to reconstruct the 3D shape from the multi-view images.
To validate the necessity of generating \textit{vector} displacement map instead of regular \textit{scalar} displacement map, we also compare with a state-of-the-art depth prediction method, DepthAnything~\cite{depthanything}, by converting the predicted depth of the object into a \textit{scalar} DM.
We run these baseline models with official implementation and pretrained weights; except that LRM does not release the official code, so we use open-source implementation OpenLRM~\cite{openlrm} instead.
For all reconstructed shapes, we render textureless images for visualization and evaluation. For Wonder3D, Magic123, and LRM, as they generate complete objects and not VDMs, we put a square plane behind their generated shapes to make the visualization more consistent and to have a fair quantitative comparison.

\input{tables/numbers_baseline}
\input{tables/numbers_ablation}

\paragraph{Evaluation Dataset and Metrics.}
As there is no existing benchmark dataset for VDMs, we collected a dataset of 50 RGB images from the Internet and a text-to-image model~\cite{adobefirefly} for evaluation. All images depict common VDM categories used by artists such as facial elements and decorations. For quantitative evaluation, we measure CLIP similarity~\cite{CLIP} and 3D-FID score~\cite{prolificdreamer} between the input image and the rendered images of the generated shapes from different views, denoted as \textbf{CLIPImg} and \textbf{3D-FID}, respectively. For CLIP, we additionally assess semantic alignment by measuring CLIP similarity between rendered images and texts describing the categories of the input images, denoted as \textbf{CLIPText}. We use public implementation of CLIP~\cite{sun2023clip} and 3D-FID~\cite{Seitzer2020FID} for computing the metrics. 

\input{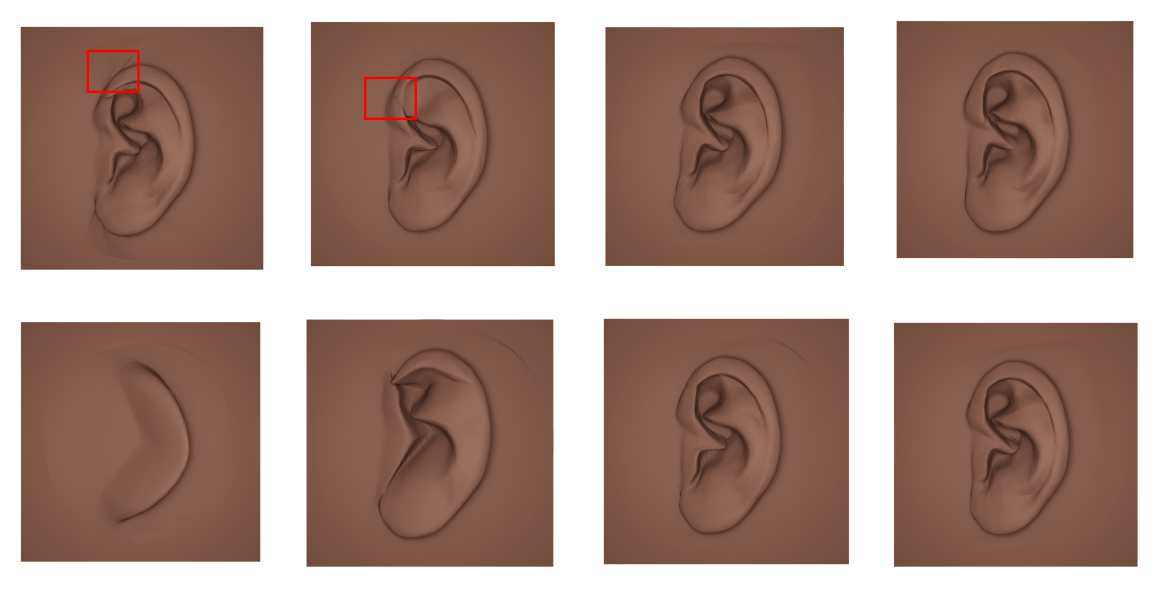}
\input{figs/ablation}

The quantitative results are summarized in Table~\ref{table:comparison} and qualitative results are presented in Figure~\ref{fig:qualitative_comparison}.
Quantitatively, our method outperforms others by a significant margin. The closest competitors to our method are Wonder3D and scalar DM, which is also reflected in the qualitative results in Figure~\ref{fig:qualitative_comparison}. Magic123 and LRM lack geometric detail as they rely heavily on textures which often hallucinate details in geometry. Wonder3D has a shape generation pipeline similar to ours, yet it was designed to generate complete objects. Therefore, it struggles to generate partial shapes, e.g., noses and ears. Although the results of scalar DM look reasonable from the front view, its side view suffers as scalar DM cannot represent unseen regions of the front view.

\vspace{-2mm}
\subsection{Ablation Study}
\label{sec:exp-ablation}

As discussed in Section~\ref{sec:vdm-recon}, we compare the following settings for parameterizing the reconstructed mesh into a VDM image: (a) Topology fixing and Tutte embedding, (b) fitting a square mesh into reconstructed mesh, and (c) our approach; see Figure~\ref{fig:vdm_fitting}. We also include the reconstructed mesh before parameterization as a reference baseline. Table~\ref{table:ablation} summarizes the quantitative results and Figure~\ref{fig:qualitative_ablation} shows the qualitative comparisons. Topological fixing and Tutte embedding suffer when the topology of the reconstructed mesh is complex due to noisy reconstruction results, as shown in Figure~\ref{fig:qualitative_ablation} (a). This is because the topological fixing algorithm does not consider the distortion after parameterization as one of its optimization goals, thus some topological fixes may significantly increase distortion. Figure~\ref{fig:qualitative_ablation} (b) shows that mesh optimization is not reliable in our setting and is likely to fall into local minima during optimization. In contrast, our method, shown in Figure~\ref{fig:qualitative_ablation} (c), not only reconstructs high quality VDMs with correct topology, but also smooths out noise induced in neural SDF reconstruction, leading to visually more pleasing results.

\input{figs/edition}
\input{figs/failcase}

\subsection{Application}
\label{sec:exp-application}

\paragraph{Shape modeling.}
With our method, users can generate parts of the shape from single-view images or text prompts (via text-to-image to obtain input to our method). Compared with methods that generate complete shapes, our method naturally provides more controllability, as users can start with a coarse shape and add customization details and shape parts, see Figure~\ref{fig:teaser}. We also show a video in the \SupplementaryMaterial to demonstrate the modeling process with VDMs generated by our method.

\paragraph{Part editing.}
With our image-to-VDM, one can perform editing in 2D image space and change the appearance of the part in 3D, see Figure~\ref{fig:edition}. Editing in image space is typically much more convenient than sculpting 3D shapes, therefore allowing users to customize their parts with ease.

%% file: tables/numbers_baseline.tex
\begin{table}[t]
\centering
\resizebox{0.999\columnwidth}{!}{
\begin{tabular}{l c c c}
\hline
\textbf{Method} & \textbf{CLIPImg↑} & \textbf{CLIPText↑} & \textbf{3D-FID↓} \\
\hline
Wonder3D~\cite{wonder3d}& 0.8246& 0.2542 & 199.5  \\
Magic123~\cite{magic123}& 0.8293& 0.2510 & 213.2  \\
LRM~\cite{LRM}& 0.8144& 0.2510 & 239.9 \\
Scalar DM&  0.8223& 0.2564 & 213.0  \\
\textbf{Ours}&  \textbf{0.8520} &\textbf{0.2701} & \textbf{192.7}  \\
\hline
\end{tabular}
}
\caption{Quantitative comparison with baselines. Scalar DM is scalar displacement map produced from DepthAnything~\cite{depthanything}.}
\label{table:comparison}
\end{table}

%% file: tables/numbers_ablation.tex
\begin{table}[t]
\centering
\resizebox{0.999\columnwidth}{!}{
\begin{tabular}{l c c c}
\hline
\textbf{Method} & \textbf{CLIPImg↑} & \textbf{CLIPText↑} & \textbf{3D-FID↓} \\
\hline

Recon. Mesh& 0.8440& 0.2636 & 198.0  \\
Topo. Fix(a)& 0.8401& 0.2617 & 209.9 \\
Mesh Opt.(b)&  0.8245& 0.2525 & 217.2  \\
\textbf{Ours(c)}&  \textbf{0.8521} &\textbf{0.2701} & \textbf{192.7}  \\
\hline
\end{tabular}
}
\caption{Quantitative ablation on VDM Reconstruction. }
\label{table:ablation}
\end{table}

%% file: figs/comparison.tex
\begin{figure*}[h]
\centering
\includegraphics[width=1.0\linewidth]{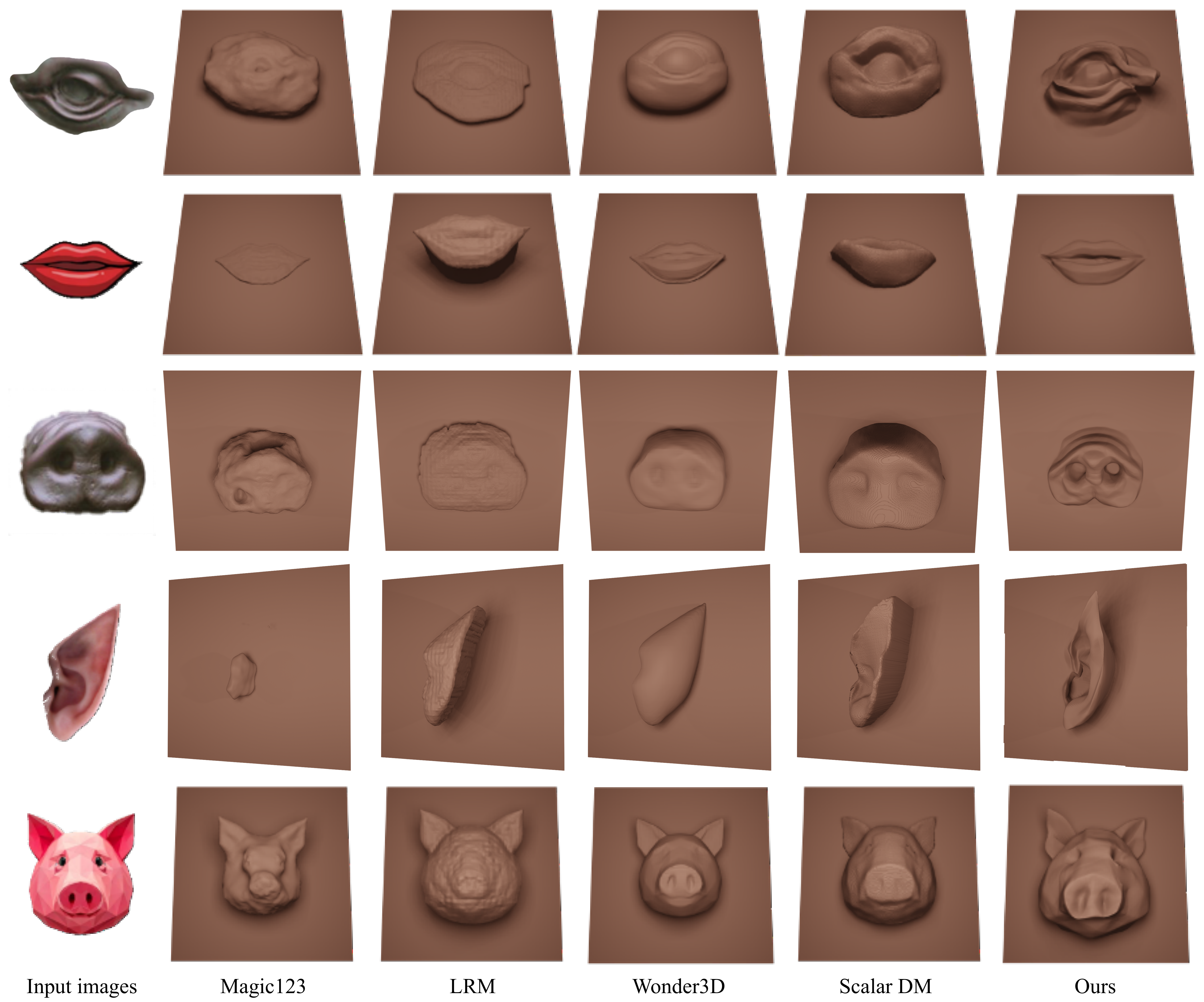}
\caption{Qualitative results compared with baseline methods. As Magic123~\cite{magic123}, LRM~\cite{LRM}, and Wonder3D~\cite{wonder3d} generate complete objects and not VDMs, we put a square plane behind their generated shapes to make the visualization more consistent.}
\label{fig:qualitative_comparison}
\end{figure*}

%% file: figs/ablation.tex
\begin{figure*}[t]
\centering
\includegraphics[width=0.85\linewidth]{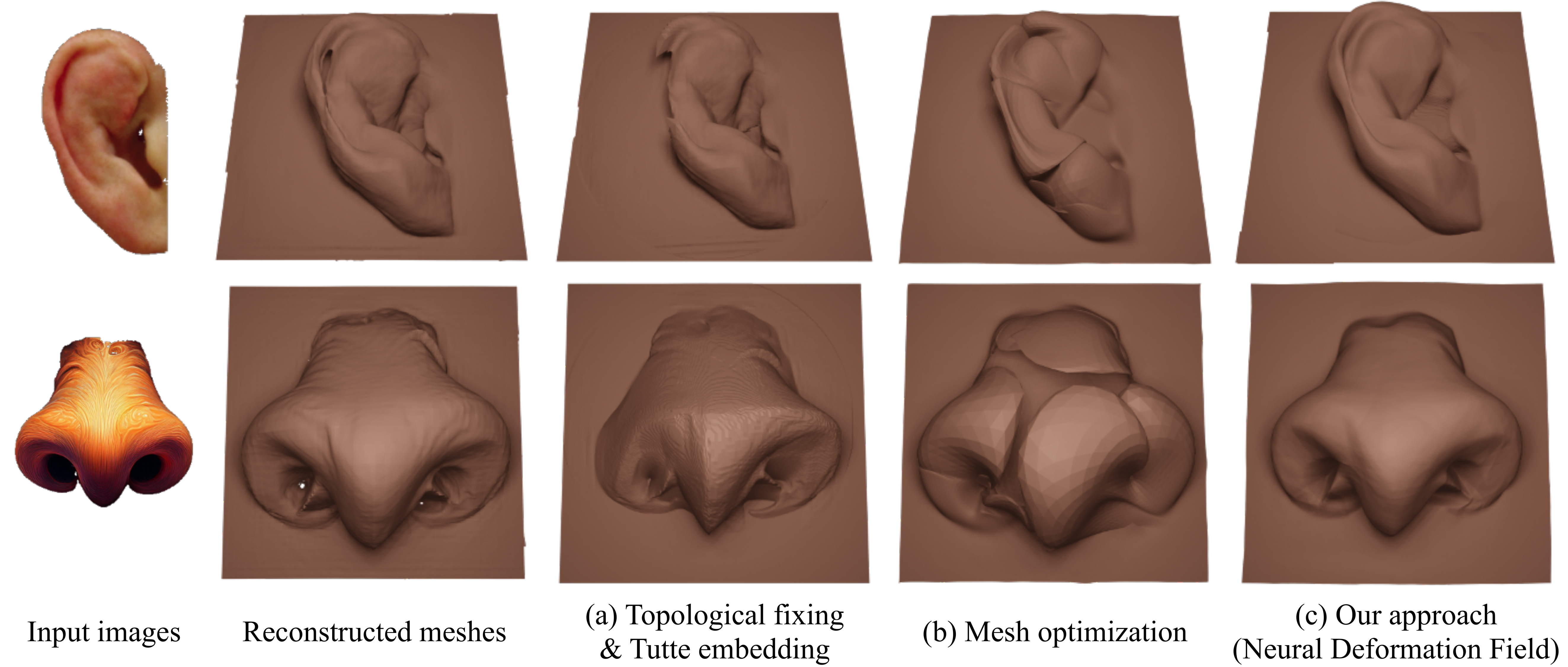}
\caption{Qualitative results of ablation study.}
\label{fig:qualitative_ablation}
\end{figure*}

%% file: figs/edition.tex
\begin{figure}[t]
  \centering
   \includegraphics[width=1.0\linewidth]{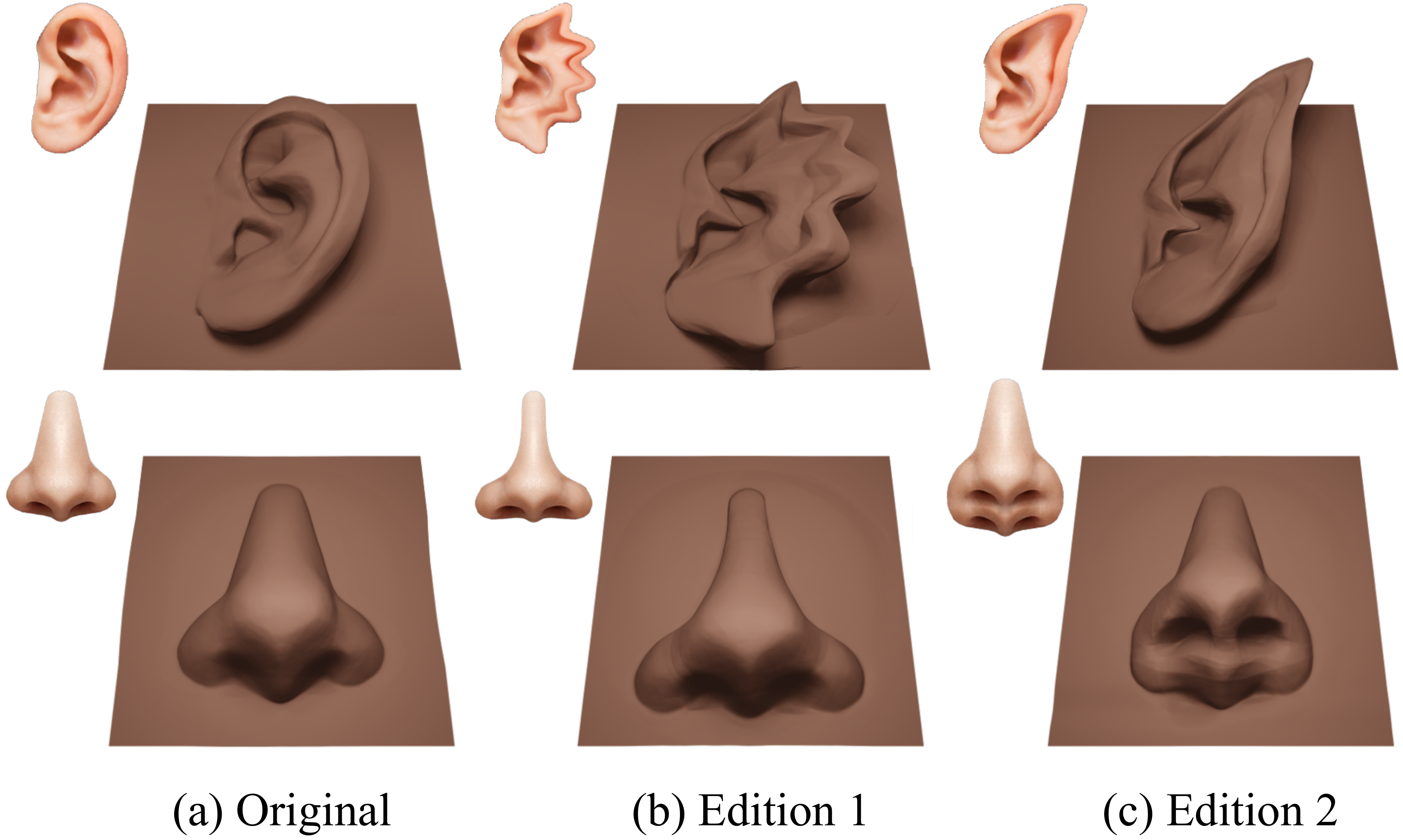}
   \caption{Customizing VDMs by editing images. Here we show original input images and generated VDMs in (a) and edited images and their generated VDMs in (b)(c).}
   \label{fig:edition}
\end{figure}

%% file: figs/failcase.tex
\begin{figure}[t]
  \centering
   \includegraphics[width=1.0\linewidth]{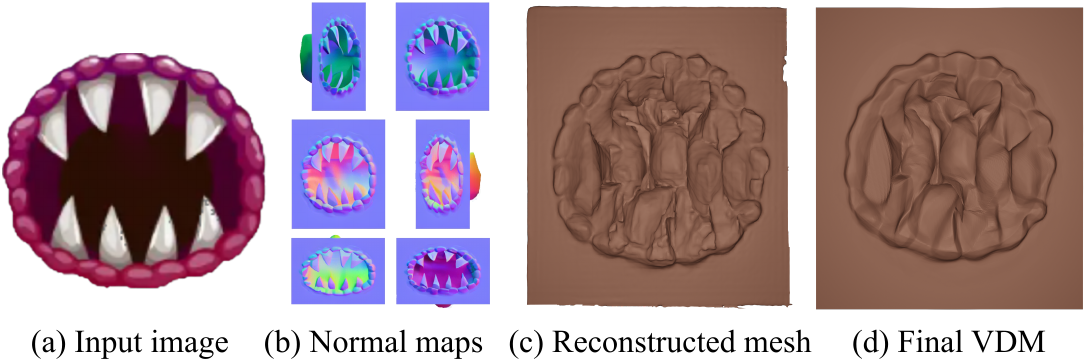}
   \caption{Failure case.}
   \label{fig:failcase}
\end{figure}

%% file: sec/5_conclusion.tex
\section{Conclusion, Limitation, and Future Work}
\label{sec:conclusions}

In this work, we propose a method to generate a VDM from an input single-view image. Our method first finetunes a pretrained image diffusion model to generate multi-view normal maps from the input image, and then reconstructs a VDM image from the multi-view normals. The generated VDMs can be used directly in shape modeling, which provide more freedom to the users on the appearance and position of each part on the shape. We also propose an efficient pipeline for creating a VDM dataset from 3D objects. Our method outperforms state-of-the-art image-to-3D models and scalar displacement map baseline, proving that our approach is more suited for VDM generation.

As discussed in Section~\ref{sec:vdm-recon}, our VDM reconstruction involves per-shape optimization, making its inference time significantly slower than the current image-to-3D methods with feed-forward LRM. Investigating the possibility of a VDM-LRM with limited training data is of great interest to us. For certain shapes with thin structures, our method cannot produce plausible results, while the generated normals look reasonable, see Figure~\ref{fig:failcase}. We suspect it is due to the multi-view images being inconsistent across different views, as observed by many other works~\cite{Vfusion3d, Sv3d}.

VDMs are predominantly used for modeling organic shapes, yet the idea of modeling-by-parts can be applied to the majority of 3D shapes. There are exciting further avenues for part-based 3D generative models.

\noindent\textbf{Acknowledgment.} Yuezhi Yang and Qixing Huang were supported by NSF IIS-2047677, NSF IIS-2413161, and Gifts from Adobe and Google.

%% file: sec/X_suppl.tex
\clearpage
\setcounter{page}{1}
\maketitlesupplementary

\appendix
\section{Details on Evaluation}
\vspace{-2mm}
For each VDM or Scalar DM, we apply it on a plane mesh and ensure that they are roughly of equal size with other comparison baselines. We add a plane behind each shape generated by single image to 3D reconstruction methods to ensure fair comparison. We render the resulting shapes in 13 different camera poses with $($elevation angle, azimuth angle$) = (0^{\circ}, \pm60^{\circ})$, $(0^{\circ}, \pm45^{\circ})$, $(0^{\circ},\pm30^{\circ})$,  $(\pm60^{\circ},0^{\circ})$, $(\pm45^{\circ},0^{\circ})$, $(\pm30^{\circ},0^{\circ})$, $(0^{\circ},0^{\circ})$ respectively. We use the default texture-less gray shading to render the shapes. For CLIP-similarity metric, we use ViT-B/32 model for evaluation. For 3D-FID score , we calculate the score between the set of rendered images and the set of input images for all shapes and use model checkpoint provided by \cite{Seitzer2020FID}. We convert the input images into gray-scale images and add a gray square behind each input image to make its appearance align with that of the rendered images. See figure \ref{fig:evalImage} for some example images used for evaluation.

\vspace{-2mm}
\section{Details on Data Preparation}
\vspace{-2mm}
As shown in figure \ref{fig:3DLasso}, Our 3D lasso tool is built upon voxel renderer. During annotation, we first select a few keypoint voxels to form a sparse loop around the region of interest, and then we find the dense loop by finding a shortest voxel path on the voxel surface that connects these selected keypoint voxels. To extract the segmented part, we remove voxels of the dense loop and use a flooding algorithm to identify the region enclosed by the annotated voxel loop. We then sample densely on the sub-mesh that is contained by the selected voxel region to obtain point clouds with normals for surface reconstruction. Since the sub-mesh may contain triangles that are not on the surface of the shape, we use fast winding number \cite{FastWindingNumber,libigl} to remove interior points. We then use Screened Poisson Surface Reconstruction implemented in Open3D\cite{Zhou2018Open3D} to remesh and obtain a single connected surface. We also filter out VDM shapes of poor quality after the data preparation pipeline to enhance data quality.
\input{figs/3DLasso}

\vspace{-2mm}
\section{Details on VDM Reconstruction}
\vspace{-2mm}
Our neural deformation field network consists of an 8-layer MLP of latent dimension 512 with residual connection at the fourth layer. We use LeakyReLU\cite{LeakyReLU} as activation function and set negative slope to 0.01. We use Adam\cite{kingma2014adam} optimizer and set learning rate to 5e-4 to optimize 3000 epochs. Specifically, we first initialize the MLP so that the initial output points form a 3D square plane. We achieve this by optimizing against an initialization optimization objective:
\[\argmin_{\theta} \;\; \frac{1}{|P|}\sum_{p \in P} \|\phi_{\theta}(p) - \text{proj}(p)\|_2^2\]
where $P$ are sets of sampled points from $[0,1]^2$ and proj$(p)$ maps p to the corresponding 3D point in a pre-defined 3D square plane. We then use the optimization objective proposed in \ref{sec:vdm-recon} for subsequent optimization. For mesh optimization comparison method, we set the laplacian regularization loss ratio to 1e-4 compared with chamfer reconstruction loss. We optimize for 1200 iterations and no remeshing is done during the optimization process. For topological fixing and tutte embedding comparison method, we use the built-in parametrization algorithm in Blender\cite{Blender} "Unwrap" function in UV editing after fixing boundary vertices. 
\vspace{-2mm}
\section{Details on Training}
\vspace{-2mm}
We finetune our multi-view normal generation model on the checkpoint provided by Zero123++\cite{zero123++} on 8 NVIDIA A100 GPUs for 3 days. We finetune it with a base learning rate of 1e-5 and dropout condition probability of 0.1. We set the batch size to 48 and optimized for 50000 steps. We do not use gradient accumulation.

\vspace{-2mm}
\section{Additional Evaluation}
\vspace{-2mm}
We added 50 RGB images from the Internet and a text-to-image model~\cite{adobefirefly} into our testing dataset of 50 images used in the paper. Quantitative results on the new 100-image dataset are shown in Table \ref{table:addi_comparison}.
\input{tables/additional_baseline}

\vspace{-2mm}
\section{More Result}
\vspace{-2mm}
We present more results generated by our model for further qualitative evaluation.

\input{figs/EvaluationImage}
\input{figs/moreResult}

%% file: figs/3DLasso.tex
\begin{figure}[t]
  \centering
   \includegraphics[width=1.0\linewidth]{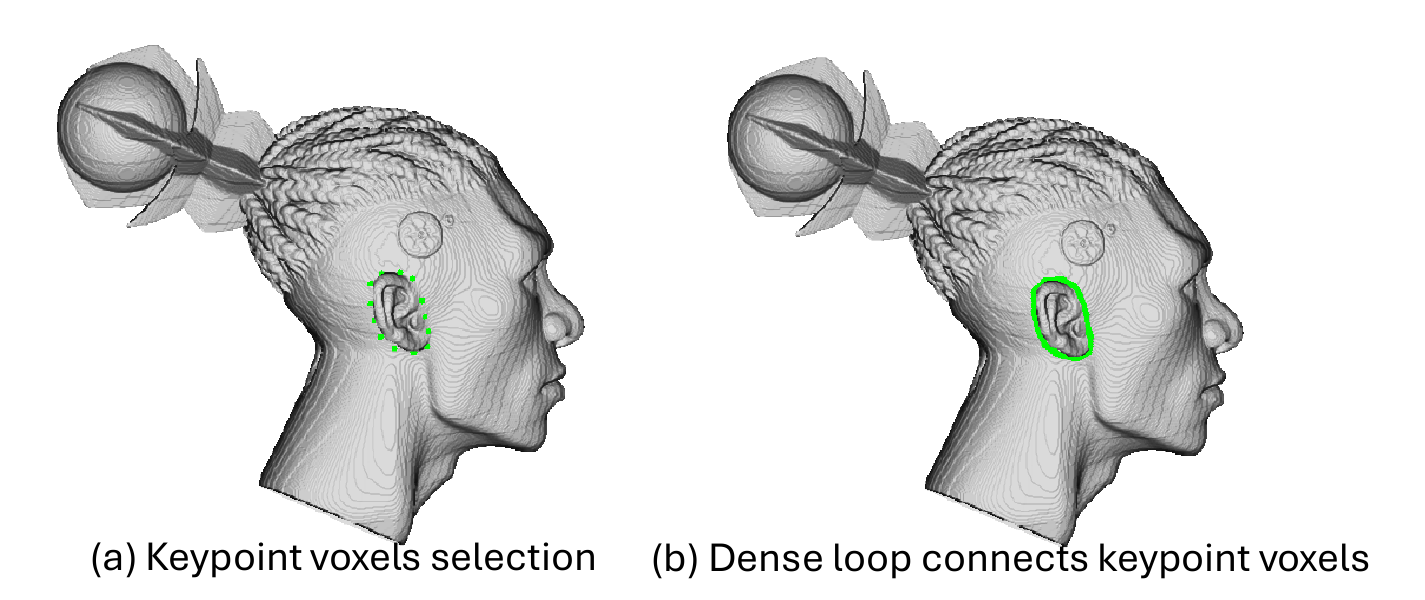}
   \caption{Our 3D lasso tool for segmentation. (a) We first select keypoint voxels around interested region. (b) We then find dense voxel loop by connecting keypoint voxels by shortest path.}
   \label{fig:3DLasso}
\end{figure}

%% file: tables/additional_baseline.tex
\begin{table}[t]
\centering
\resizebox{0.999\columnwidth}{!}{
\begin{tabular}{l c c c}
\hline
\textbf{Method} & \textbf{CLIPImg↑} & \textbf{CLIPText↑} & \textbf{3D-FID↓} \\
\hline
Wonder3D~\cite{wonder3d}& 0.8143& 0.2504 & 208.7  \\
Magic123~\cite{magic123}& 0.8241& 0.2488 & 218.5  \\
LRM~\cite{LRM}& 0.8122& 0.2472 & 246.3 \\
Scalar DM&  0.8201& 0.2529 & 223.4  \\
\textbf{Ours}&  \textbf{0.8460} &\textbf{0.2639} & \textbf{197.3}  \\
\hline
\end{tabular}
}
\caption{Additional quantitative comparison with baselines.}
\label{table:addi_comparison}
\end{table}

%% file: figs/EvaluationImage.tex
\begin{figure*}[h]
  \centering
   \includegraphics[width=0.7\linewidth]{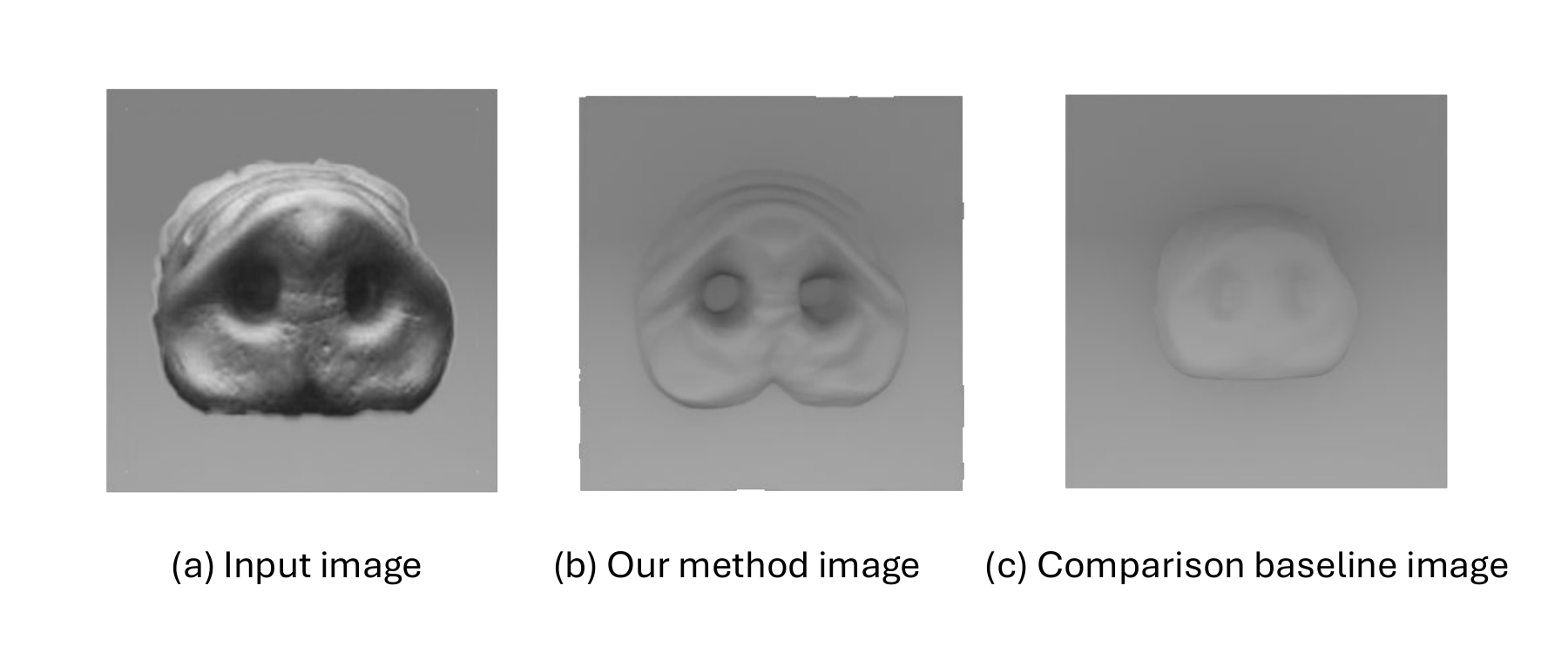}
   \caption{Example images used for computing quantitative results. (a) Input image. (b) Rendered image of our method. (c) Rendered image of a comparison baseline method, Wonder3D\cite{wonder3d}.}
   \label{fig:evalImage}
\end{figure*}

%% file: figs/moreResult.tex
\begin{figure*}[t]
  \centering
   \includegraphics[width=1.0\linewidth]{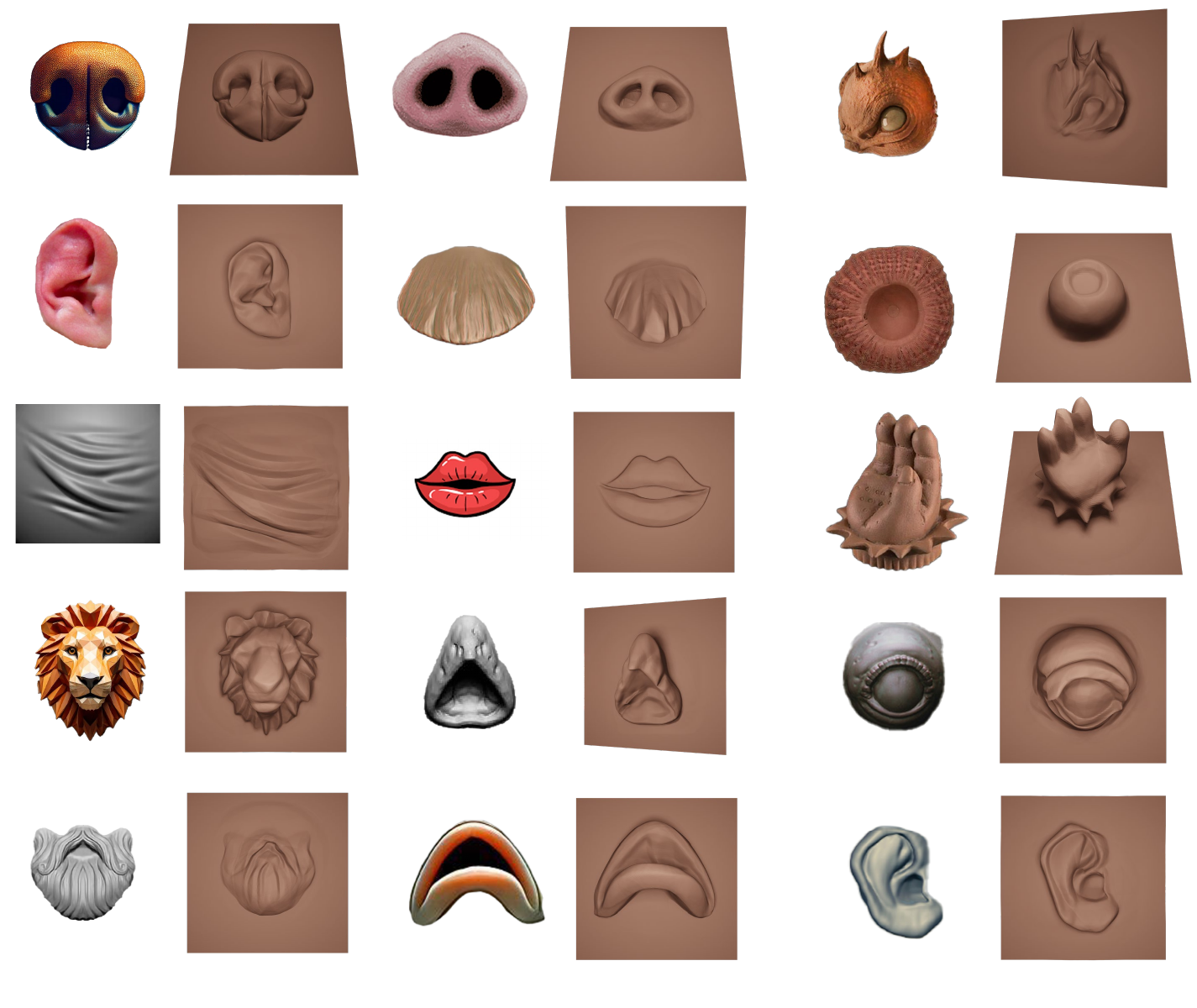}
   \caption{More results.}
   \label{fig:moreresult}
\end{figure*}